\documentclass[pre,preprint,amsmath,amssymb,aps,nofootinbib]{revtex4-1}
\usepackage{amssymb,amsmath}
\usepackage{graphicx}

\begin{document}

\title{Studying of nonlinear normal modes interactions in $SF_6$ molecule with the aid of the density functional theory}

\author{G. Chechin}
 \email{gchechin@gmail.com}
\author{D. Ryabov}
\author{S. Shcherbinin}
\affiliation{Research Institute of Physics, Southern Federal University, 194 Stachki Ave., Rostov-on-Don 344090, Russia}

\begin{abstract}
Some exact interactions between vibrational modes in systems with discrete symmetry can be described by the theory of the bushes of nonlinear normal modes (NNMs) [G.M.~Chechin, V.P.~Sakhnenko.\ Physica~D 117, 43 (1998)]. Each bush represents a dynamical object conserving the energy of the initial excitation. Existence of bushes of NNMs is ensured by some group-theoretical selection rules.

In [G.M.~Chechin, \emph{et~al.} Int.\ J.\ Non-Linear Mech.\ 38, 1451 (2003)], existence and stability of the bushes of vibrational modes in the simple octahedral model of mass points interacting via Lennard-Jones potential were investigated. In the present paper, we study these dynamical objects by the density functional theory in $SF_6$ molecule which possesses the same symmetry and structure. We have fully confirmed the results previously obtained in the framework of the group theoretical approach and have found some new properties of the bushes of NNMs.
\end{abstract}

\maketitle

\section{Introduction}
\label{Intro}
We consider nonlinear vibrations in Hamiltonian systems. Conventional, or linear normal modes (LNMs) are exact solutions of dynamical equations in the harmonic approximation~\cite{landau}. This approximation means, that potential energy of the considered system is decomposed into multi-dimensional Taylor series, and only quadratic terms are taken into account for obtaining Newton classical equations. LNMs cease to be exact solutions when we involve not only quadratic, but also some anharmonic terms. If the latter terms are sufficiently small, then it is possible to speak about some interactions between LNMs. In many physical problems, such interactions and the role of different anharmonic terms are studied. Taking into account the smallness of these terms, we can construct some \emph{approximate} solutions for the considered nonlinear problem. As an example let us refer to studying the various types of the phonon-phonon interactions in crystal physics~\cite{Ziman}.

In this connection one can ask, are there exist some \emph{exact} solutions of nonlinear dynamical equations beyond harmonic approximation?

Let we have a nonlinear system whose potential energy contains besides quadratic terms (therefore, the system admits the harmonic approximation) some anharmonic terms preceded by a common factor $\gamma$. This parameter determines the strength of nonlinearity of the system. Lyapunov proved~\cite{Lyapunov} that each LNM can be continued in $\gamma$ to obtain an exact periodic solution of the nonlinear system. He also gave a procedure to construct this exact solution up to arbitrary degree of the parameter $\gamma$. As a result, one can find the set of $N$ Lyapunov nonlinear normal modes (NNMs), where $N$ is the dimension of the system. Unfortunately, the procedure of Lyapunov NNMs constructing is very cumbersome, and we can obtain solutions only up to a small degree of the parameter $\gamma$. Moreover, this procedure usually converges only for very small values $\gamma$. These factors prevent to use the Lyapunov modes in many physical problems.

Another notion of nonlinear normal modes was introduced by Rosenberg~\cite{rozenberg} (see also~\cite{vakakis}). These modes can exist for large values of the parameter $\gamma$ and even for the essentially nonlinear systems whose potential energy does not contain any quadratic terms (such systems don't admit the harmonic approximation). The number of NNMs by Rosenberg, if they exist, can be less or even greater than the full number of the system's degrees of freedom.

Unfortunately, Rosenberg modes can exist only in very specific nonlinear systems. Let us consider this point in more detail for N-degrees-of-freedom mechanical system. According to definition, in the dynamical regime described by a given Rosenberg mode, all degrees of freedom, $x_i(t)$, vibrate as follows:
\begin{equation}\label{NNM}
x_i(t)=a_i f(t),~i=1..N.
\end{equation}
Thus, all $x_i(t)$ at any moment t are proportional to the \emph{same} time-dependent function $f(t)$ ($a_i$ are constant coefficients). Actually, this means a separation of space and time variables\footnote[1]{The LNMs also satisfy Eq.~\eqref{NNM} with $f(t)=a\cos(\omega t + \phi_0)$, where $a$, $\phi_0$ and $\omega$ are amplitudes, initial phase and frequency, respectively.}.

Substitution of the ansatz~\eqref{NNM} into differential equations of the dynamical system leads to the set of $(N-1)$ algebraic equations for obtaining amplitudes $a_i$ and to one differential equation which determines time-periodic function $f(t)$. The latter equation is called ``governing''.

Rosenberg found several classes of mechanical systems for which Eq.~\eqref{NNM} is fulfilled. The most important class is formed by systems whose potential energy is a homogeneous function, of arbitrary degree, of all its arguments.

In~\cite{DAN-1, DAN-2}, we have shown that existence of Rosenberg nonlinear normal modes can be a direct consequence of a certain discrete symmetry of the considered physical system. We refer to such modes as symmetry-determined nonlinear normal modes (SD-NNMs). Hereafter only this type of Rosenberg modes is considered.

In~\cite{DAN-1, DAN-2}, we have also introduced the concept of \emph{bushes of NNMs} for dynamical systems described by arbitrary group $G_0$ of discrete symmetry. Every bush possesses a certain symmetry group $G$ which is a subgroup of the group $G_0$ ($G\subset G_0$) and represents an \emph{exact} solution of nonlinear dynamical equations. As a consequence, the energy of initial excitation of a given bush turns out to be trapped in this dynamical object until it loses its stability (then the bush transforms into another bush of lower symmetry and larger dimension). Every bush whose dimension $m$ is greater than unity ($m>1$) describes a \emph{quasiperiodic} dynamical regime which is determined by $m$ governing differential equations. In this sense, one can consider the Rosenberg mode as one-dimensional bush.

The specific group-theoretical methods for constructing bushes of NNMs were developed in~\cite{DAN-1, DAN-2, PHD-1998}. It is essential that because of symmetry-related causes only a finite number of the bushes of each dimension can exist in the physical system with a given symmetry group~$G_0$. For example, it was found~\cite{FPU-1, FPU-2, scsh} that for monoatomic chain with even interparticle interactions and periodic boundary conditions (such as FPU-$\beta$ chain or electrical chain, considered in~\cite{Osting}) only five SD-NNMs can exist (depending on the number of particles in the chain). All bushes of low dimensions for many mechanical structures with various types of translational and point symmetry were found in~\cite{DAN-1, DAN-2, CSSSH, octo}. In particular, in~\cite{CSSSH}, all SD-NNMs (one-dimensional bushes) were found for all possible mechanical structures with any of $230$ space symmetry groups.

Note that construction of bushes can be done \emph{without} any information about the type of interparticle interactions in the given physical system. One must know only its symmetry group and the geometrical structure.

Searching for the complete set of modes entering the bush with a given symmetry, we deal with its geometrical aspect. This problem can be solved with the aid of group-theoretical method only. On the other hand, we deal with dynamical aspect of the bush when study time-evolution of its modes and the bush stability. In this stage of bush studying, we must know specific interactions between particles of the physical system.

In~\cite{octo}, we have studied geometrical and dynamical properties of one-, two- and three-dimensional bushes in a simple octahedral mechanical system which is depicted in Fig.~\ref{fig1}. It represents a regular octahedron with equal point masses at the vertices and with another point mass in its center. It was found, that $18$ bushes of vibrational modes can exist in the case when the central particle is immovable. Among these bushes there are $1D$, $2D$ and $3D$ bushes with point symmetry groups $O_h$, $D_{4h}$ and $C_{4v}$, respectively. These bushes were studied in~\cite{octo} not only as geometrical objects, but also as dynamical objects supposing that interparticle interactions in the considered system are described by arbitrary pair potential, $U(r)$, in particular, by the Lennard-Jones potential
\begin{equation}
U(r)=\frac{A}{r^{12}} - \frac{B}{r^6}.
\end{equation}

Naturally, one can ask: ``Are there exist some physical systems whose nonlinear dynamics can be described by the above simple mechanical model?'' The more general question can be formulated as follows: ``Can the concept of bushes of nonlinear normal modes and the methods of their studying be valid for nonlinear dynamics of \emph{real physical systems}?''

First of all, let us note that there are some molecules, whose equilibrium state corresponds to the mechanical model in Fig.~\ref{fig1}. As an example, we choose the molecule $SF_6$. The experimental investigation of nonlinear vibrations of such molecules represents great difficulties (see, for example, review paper~\cite{itikawa} devoted to study interactions between normal modes in simple molecules).

On the other hand, the ab initio calculations based on the density functional theory (DFT)~\cite{Kohn-Hohenberg, Kohn-Sham, Kohn-Nob}, proved to be very effective and rather correct for studying molecules and crystals. Indeed, DFT allows to determine the geometrical structures of many microscopic objects up to $1\%$ accuracy~\cite{Kohn-Nob}. Therefore, it is very interesting to study nonlinear dynamics in the framework of DFT methods. In our recent paper~\cite{Dmitriev}, we successfully realized such a program for studying discrete breathers in graphane. We found there that DFT leads to cardinally different results for vibrations with large amplitudes than the methods of molecular dynamics based on the concept of mass points interacting via a widely used phenomenological pair potential. Indeed, the mass-point models cannot describe the complicated process of the polarisation of the atomic shell during essentially nonlinear vibration. In~\cite{Dmitriev1}, the Brenner potential~\cite{brenner} was used for studying discrete breathers and the non-monotonic connection $\nu(A)$ between breather amplitude $A$ and its frequency $\nu$ was obtained (it shows two changes between soft and hard nonlinearity). In contrast, with the aid of ABINIT package~\cite{ABINIT}, we have obtained the monotonic function $\nu(A)$, which demonstrates that only soft nonlinearity takes place for gap breathers in graphane.

In this paper, we study nonlinear dynamics of $SF_6$ by DFT methods realized in ABINIT code~\cite{ABINIT}. The paper is organized as follows. In Sec.~2, the geometrical aspects of nonlinear vibrations of $SF_6$ corresponding to the bushes with point symmetry groups $O_h$, $D_{4h}$, $C_{4v}$ are discussed. Sec.~3 is devoted to ab initio calculations of the dynamics of the above bushes. In Sec.~4, we comment on the validity of the bushes of the nonlinear normal modes in real nonlinear systems (using $SF_6$ molecule as an example) and discuss further possible investigations of nonlinear dynamics of physical systems in the framework of DFT.

\section{Group-theoretical analysis of nonlinear vibrations in $SF_6$}\label{sec:2}

\begin{figure}[htb]
\begin{center}
\includegraphics[scale=0.5]{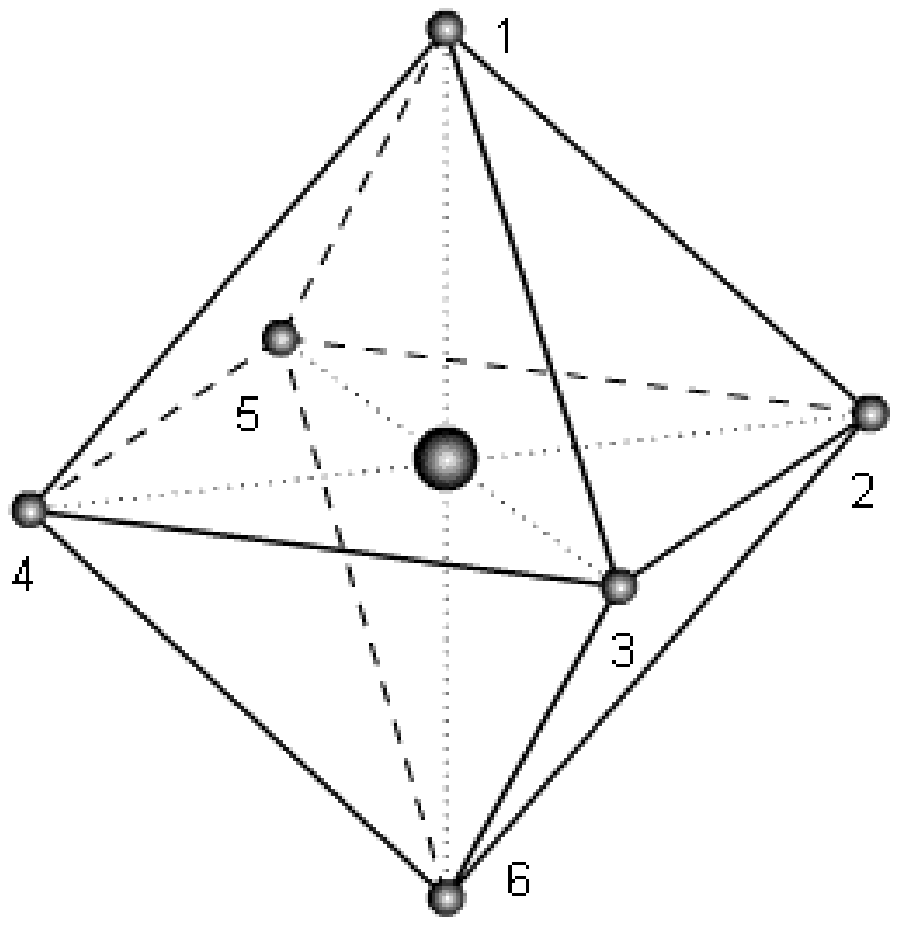}
\caption{Octahedral molecule.}\label{fig1}
\end{center}
\end{figure}

Group-theoretical analysis of nonlinear vibrations of the structure presented in Fig.~\ref{fig1} was already fulfilled in Ref.~\cite{octo}. Here we reproduce some results of that work which are necessary for the further discussion.

In any vibrational regime, the nuclei of the atoms, consisting the $SF_6$ molecule, displace from the equilibrium positions depicted in Fig.~\ref{fig1} and we can speak about a displacement pattern at any fixed moment $t$. For every NNM, as well as for LNM, this pattern possesses a certain point symmetry group. In the conventional phonon-spectrum analysis (this is the case of small vibrations), the above patterns are determined by eigenvectors of the matrix of force constants.

There exists Wigner theorem~\cite{Wigner} about group-theoretical classification of the linear normal modes. According to this theorem, the modes are classified by \emph{irreducible representations} (irreps) of the symmetry group $G_0$ of the system in equilibrium. In this way, we can introduce the basis $\boldsymbol{\Phi}=\{\boldsymbol{\phi_j}| j=1..N\}$ in the space of all possible atomic displacements, constituted by the complete set of basis vectors of the irreps entering into the mechanical representation of the considered system. Therefore, any vibrational regime $\textbf{X}(t)=\{x_1(t), x_2(t),\ldots,x_n(t)\}$ in this system can be decomposed into the above basis with coefficients depending on time~$t$:
\begin{equation}\label{raz-1}
\textbf{X}(t)=\sum_{j=1}^Nc_j(t)\boldsymbol{\phi_j}\equiv(C(t), \boldsymbol{\Phi}).
\end{equation}
In this equation, each term $c_j(t) \boldsymbol{\phi}_j$ can be considered as nonlinear normal mode according to the definition~\eqref{NNM}. Indeed, the vector multiplier $\boldsymbol{\phi_j}$ determines the displacement pattern of all atoms, i.e.\ the space structure of NNM, while $c_j(t)$ determines time-evolution of the mode. However, for brevity, we often use the term nonlinear normal mode (or vibrational mode) individually for $\boldsymbol{\phi_j}$, as well as for $c_j(t)$.

The basis vectors $\boldsymbol{\phi_j}$ correspond to different irreps $\Gamma_n$ of the group $G_0$ and, therefore, the displacement vector $\textbf{X}(t)$ in~\eqref{raz-1} can be written as the sum of contributions associated with individual representations of the equilibrium symmetry group $G_0$:
\begin{equation}\label{raz-rep}
\textbf{X}(t)=\sum(\textbf{C}_n(t),\boldsymbol{\Phi}[\Gamma_n]).
\end{equation}
Here, $\boldsymbol{\Phi}[\Gamma_n]$ is the set of basis vectors of the irrep $\Gamma_n$.

According to Wigner theorem, the \emph{small} vibrations of the molecule associated with the different irreducible representations $\Gamma_n$ are \emph{independent from each other}. It means that if one excite (by using the appropriate initial conditions when solving linear differential equations) a dynamical regime $\textbf{X}(t)$ corresponding to a given irrep $\Gamma_n$, this regime can never leads to excitation of the modes belonging to another irreps in the decomposition~\eqref{raz-rep}. Therefore, one can ask: ``What will happen if we consider large and, therefore, \emph{nonlinear} vibrations of the molecule?'' The theory of the bushes of nonlinear normal modes starts from this question.

The answer was given in Ref.~\cite{DAN-1} (see also~\cite{PHD-1998}, devoted to discussion of the bush theory). It turns out that there exist certain \emph{selection rules} for excitation transfer from one mode to another. These rules are originate from some group-theoretical restrictions which can be written as a certain system of linear algebraic equations~\cite{DAN-1}. In particular, one can deduce from this system that excitation from the mode with the given symmetry group $G$ can transfer only to those nonlinear normal modes whose own symmetry is \emph{higher} or \emph{equal} to $G$. The above selection rules lead to possibility for existence of bushes of NNMs.

Each bush represents a set of NNMs that conserves the energy of initial excitation until it loses stability because of the phenomenon similar to the parametric resonance with some modes outside a given bush. This phenomenon occurs if amplitudes of some bush modes attain sufficiently large values (see details in~\cite{PHD-1998, FPU-2}).

Every bush possesses its own symmetry that is determined by intersection of all symmetry groups of its modes. As was already mentioned, when the given bush loses stability, it transforms into another bush with lower symmetry and with higher dimension.

Let us consider the simplest bushes for nonlinear vibrations of $SF_6$ molecule using some results obtained in~\cite{octo}.
In the equilibrium state, depicted in Fig.~\ref{fig1}, the molecule $SF_6$ possesses point symmetry group $G_0=O_h$. All vibrational modes for this molecule, classified by irreps of the group $O_h$, can be found in Table.~3 in~\cite{octo} [three translational modes must be excluded since the central atom ($S$) is supposed to be immovable].

For the present consideration, we need explicit forms of the displacement patterns of NNMs $\boldsymbol{\phi}_1$, $\boldsymbol{\phi}_2$, $\boldsymbol{\phi}_3$, corresponding to one-dimensional irrep $\Gamma_1$, two-dimensional irrep $\Gamma_5$ and three-dimensional irreps $\Gamma_{10}$. These displacement patterns are given in Table~\ref{table1}.

\begin{table}[h]
\caption{Displacement patterns of the NNMs in $SF_6$ molecule}
\begin{center}\label{table1}
\begin{tabular}{|p{3em}|p{3em}|p{22em}|}
  \hline
   Irrep & NNM & Pattern \\
  \hline
  $\Gamma_1$ & $\boldsymbol{\phi}_1$ & $\frac{1}{\sqrt{6}}(0,0,-1|-1,0,0|0,-1,0|1,0,0|0,1,0|0,0,1|)$ \\
  \hline
  $\Gamma_5$ & $\boldsymbol{\phi}_2$ & $\frac{1}{\sqrt{12}}(0,0,2|-1,0,0|0,-1,0|1,0,0|0,1,0|0,0,-2|)$ \\
  \hline
  $\Gamma_{10}$ & $\boldsymbol{\phi}_3$ & $\frac{1}{\sqrt{12}}(0,0,-2|0,0,1|0,0,1|0,0,1|0,0,1|0,0,-2|)$ \\
  \hline
\end{tabular}
\end{center}
\end{table}

In this table, for each fluorine atom, according to the numbering in Fig.~\ref{fig1}, we point out three coordinates $x, y, z$, which determines displacement of the nucleus of this atom from the equilibrium position. One can see from Table 1 that the molecule shape in the vibrational regime, corresponding to the mode $\boldsymbol{\phi}_1$ represents, at \emph{any} moment $t$, the regular octahedron. Its size vibrates in time becoming larger or lesser in comparison with the octahedron corresponding to the equilibrium state. This NNM is called ``breathing'' mode. It represents one-dimensional bush with symmetry group $O_h$. The symmetry group of the breathing mode, $O_h$, is higher than that of each other vibrational mode and, therefore, according to the above mentioned selection rules, the excitation from this mode cannot transfer to another modes. Therefore, the breathing mode will vibrate for arbitrary long time without involving into the dynamical regime any other vibrational modes. In other words, if we decompose $\textbf{X}(t)$ for the breathing mode in accordance with the Eq.~\eqref{raz-1}, we obtain that right-hand side of this equation is reduced to one term only
\begin{equation}\label{raz-oh}
\textbf{X}(t)=c_1(t)\boldsymbol{\phi}_1\equiv a(t)\boldsymbol{\phi}_1.
\end{equation}
Here we have renamed the time-dependent coefficient $c_1(t)$ as $a(t)$.

This means that breathing mode represents an \emph{exact} solution to the nonlinear differential equations describing dynamics of our mechanical system for \emph{any type} of interatomic interactions. The explicit form of the differential equation for the time-dependent coefficient $a(t)$ can be found if one substitute Eq.~\eqref{raz-oh} into the full original system of $3\times 6=18$ nonlinear equations (remember that the central atom is immovable) taking into account the specific type of interatomic forces. In this way, we reveal that the above system of $18$ equations is reduced to only one differential equation for $a(t)$. Let us emphasize that $a(t)$ indeed depends on interatomic forces, while the displacement pattern \emph{does not depend} on these forces and can be found with the aid of the group-theoretical method only (see Table~\ref{table1}).

Thus, the breathing mode represents one-dimensional bush with symmetry group $O_h$ and we denote it by symbol $B[O_h]$. In~\cite{octo}, one can find an explicit form of the differential equation for $a(t)$ in the case of arbitrary pair interatomic potential $U(r)$, as well as for Lennard-Jones potential.

\begin{figure}[h]
\begin{minipage}[h]{0.31\linewidth}
\center{\includegraphics[width=1\linewidth]{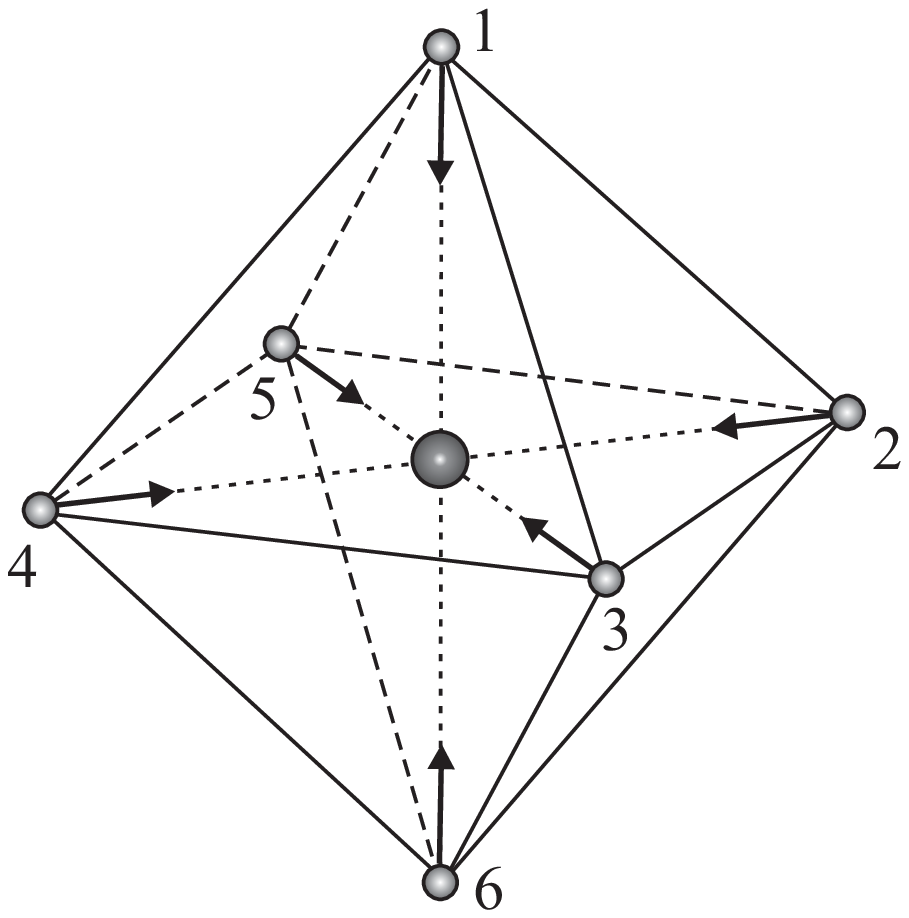} \\~\\ a)}
\end{minipage}
\hfill
\begin{minipage}[h]{0.31\linewidth}
\center{\includegraphics[width=1\linewidth]{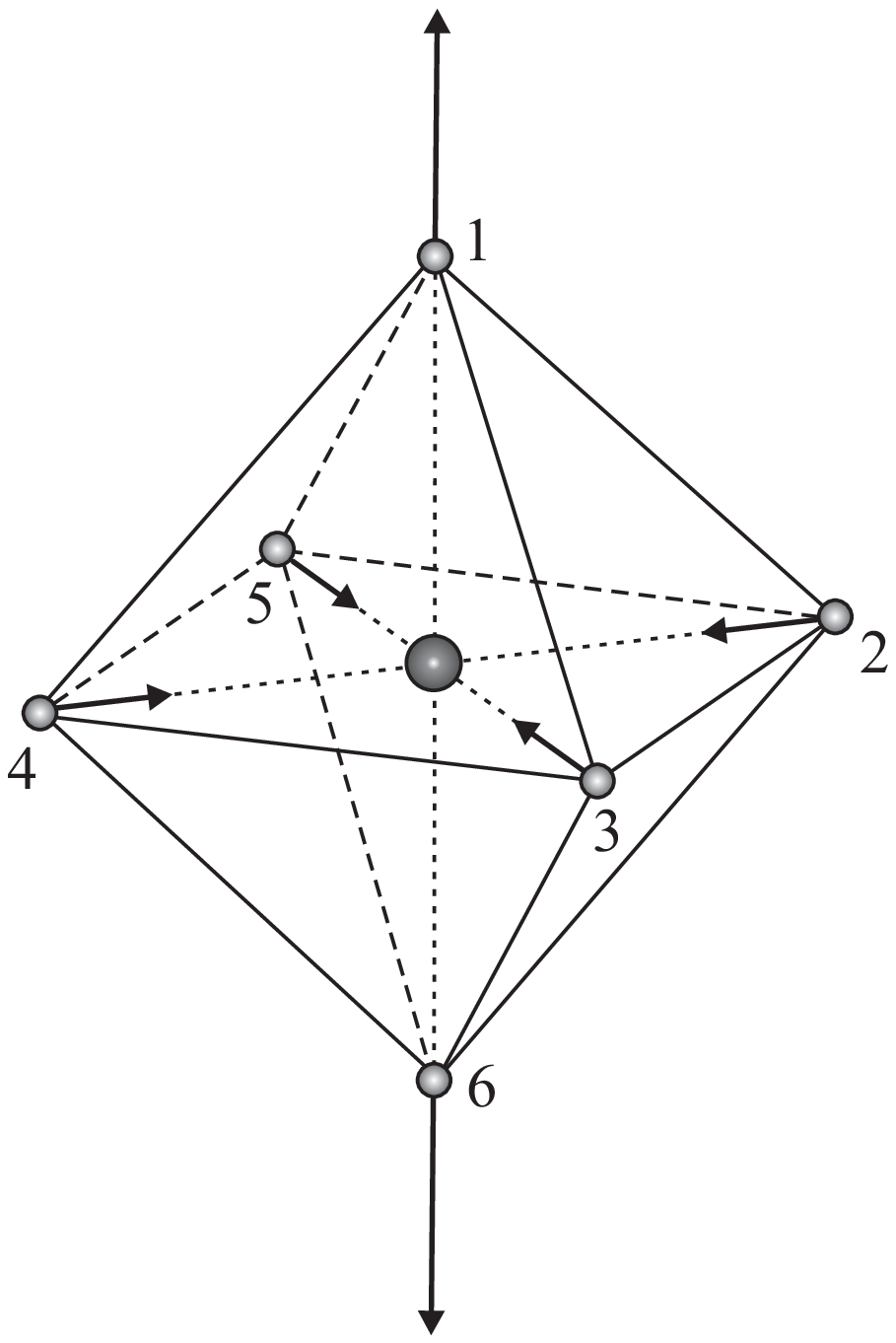} \\~\\ b)}
\end{minipage}
\hfill
\begin{minipage}[h]{0.31\linewidth}
\center{\includegraphics[width=1\linewidth]{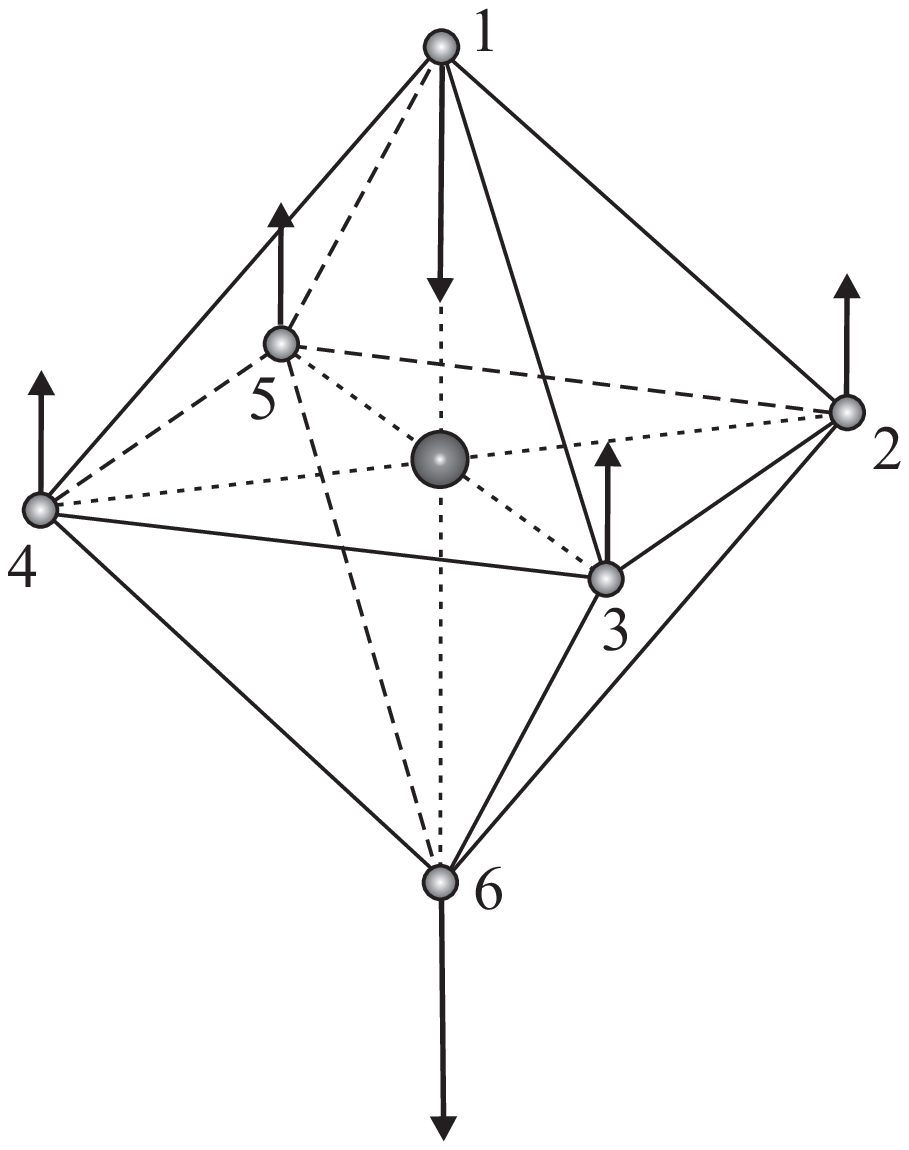} \\ c)}
\end{minipage}
\caption{Displacement patterns for modes $\boldsymbol{\phi}_1$, $\boldsymbol{\phi}_2$ and $\boldsymbol{\phi}_3$.}
\label{fig2}
\end{figure}

Another situation appears when we excite the mode $\boldsymbol{\phi}_2$ by displacing fluorine atoms according to its pattern given in Table~\ref{table1}. The own point symmetry of this mode is described by the point group $G_2=D_{4h}$ that is a subgroup of the equilibrium state symmetry group $G_0=O_h$ ($D_{4h} \subset O_h$). Displacement pattern associated with $\boldsymbol{\phi}_2$ represents octahedron with square base, formed by atoms $2$, $3$, $4$, $5$ (see Fig.~\ref{fig1}), and the atoms $1$, $6$ situated on $Z$ axis and displaced by the same distance from their equilibrium positions from the octahedron base. In Fig.~\ref{fig2}, we depict the above displacements of all six $F$ atoms by arrows. In contrast to the case of excitation of the breathing mode, in the present case, the mode $\boldsymbol{\phi}_2$ cannot exist independently of all other modes. Indeed, its excitation leads to excitation of the breathing mode $\boldsymbol{\phi}_1$ whose symmetry group $G_1=O_h$ is higher than the symmetry group $G_2=D_{4h}$. This conclusion is a result of the group-theoretical analysis only. Thus, we obtain the two-dimensional bush $B[D_{4h}]$ with the point group $G_2=D_{4h}$. From the general decomposition~\eqref{raz-1}, we obtain for this case:
\begin{equation}\label{ras-d4h}
\textbf{X}(t)=c_2(t)\boldsymbol{\phi}_2+c_1(t)\boldsymbol{\phi}_1\equiv b(t)\boldsymbol{\phi}_2+a(t)\boldsymbol{\phi}_1.
\end{equation}
Here we have renamed $c_1(t)$ and $c_2(t)$ by $a(t)$ and $b(t)$, respectively.

Substitution of the anzats ($6$) into $18$ original nonlinear equations confirms that all these equations are transformed into two differential equations with respect to the unknown coefficients $a(t)$ and $b(t)$. These two time-dependent functions fully describe the dynamics of the two-dimensional bush $B[D_{4h}]$\footnote{Note, that $a(t)$ in Eq.~\eqref{ras-d4h} is not in any relation with $a(t)$ from Eq.~\eqref{raz-oh}!}. It describes \emph{quasiperiodic} vibrations with two basic frequencies $\omega_1$ and $\omega_2$ corresponding to the modes $a(t)\boldsymbol{\phi}_1$ and $b(t)\boldsymbol{\phi}_2$ (because of nonlinearity of the considered system, in the Fourier spectrum of such vibrational regime, one can observe not only the frequencies $\omega_1$, $\omega_2$, but also their different integer linear combination).

Let us emphasize once more that \emph{no other modes} present in the decomposition~\eqref{raz-1}, i.e.\ Eq.~\eqref{ras-d4h} represents an \emph{exact} solution of the original nonlinear equations whose coefficients $c_1(t)=a(t)$ and $c_2(t)=b(t)$ can be obtain as solution of two governing differential equations depending on the specific interatomic interactions. In~\cite{octo}, we presented the explicit expressions of these governing equations for the case of arbitrary pair potential $U(r)$ and, in particular, for the Lennard-Jones potential.

Let us note that in the vibrational regime described by the bush $B[D_{4h}]$ the atom configuration represents the octahedron with square base and with atoms $1$ and $6$ displaced by the same distance in opposite directions (this is a consequence of the presence in the group $D_{4h}$ horizontal mirror plane determined by the base of octahedron).

The mode $\boldsymbol{\phi}_3$ (see Table~\ref{table1}), being excited at the initial instant $t=t_0$, generates \emph{three-dimensional} bush $B[C_{4v}]$ whose full symmetry coincides with the group $G_3=C_{4v}$ of the mode $\boldsymbol{\phi}_3$.

The modes whose symmetry determines the symmetry of the whole bush we call ``root modes'', while other modes which turns out to be excited automatically as a result of the root mode excitation we call ``secondary modes''. Let us note, that in general the situation can be more complicated. Indeed, even for the original group $G_0=O_h$ there exist bushes whose excitation is possible only by exciting \emph{simultaneously} several modes with different symmetry groups. In this case the symmetry of the whole bush is determined by \emph{intersection} of the above symmetry groups. All nontrivial bushes of different dimensions (up to dimension equal to $8$) were presented in~\cite{octo}.

In contrast to the group $G_2=D_{4h}$, the horizontal mirror plane is absent in the group $G_3=C_{4v}$. As a consequence, displacements of the fluorine atoms $1$ and $6$ need not to be equal in magnitude and, therefore, the octahedron corresponding to the bush $B[C_{4v}]$ possesses different heights dropped from the atoms $1$ and $6$ to the octahedron base formed by the atoms $2$, $3$, $4$,~$5$.

Thus, we obtain the three-dimensional bush $B[C_{4v}]$ to which the following decomposition of $\textbf{X}(t)$ corresponds:
\begin{equation}\label{raz-c4v}
\textbf{X}(t)=c_1(t)\boldsymbol{\phi}_1+c_2(t)\boldsymbol{\phi}_2+c_3(t)\boldsymbol{\phi}_3\equiv a(t)\boldsymbol{\phi}_1+b(t)\boldsymbol{\phi}_2+c(t)\boldsymbol{\phi}_3.
\end{equation}
Here we have renamed the coefficients $c_1(t)$, $c_2(t)$, $c_3(t)$ by $a(t)$, $b(t)$, $c(t)$, respectively. Substitution of the anzats~\eqref{raz-c4v} into the original $18$ nonlinear differential equations leads to three governing equations (all other original equations turn out to be equivalent to these governing equations) whose explicit form, for the case of pair potential $U(r)$, are given in~\cite{octo}.

For all above-discussed bushes, the symmetry of the \emph{vibrational state} conserve, as well as the complete set of the bush modes, while amplitudes of these modes [they are described by coefficients $c_i(t)$ in Eqs.~\eqref{raz-oh},\eqref{ras-d4h},\eqref{raz-c4v}] evaluate in time.

We have considered three simplest bushes $B[O_h]$, $B[D_{4h}]$, $B[C_{4v}]$ in the $SF_6$ molecule in the framework of the group-theoretical approach. The existence of these bushes was \emph{confirmed} by straightforward numerical experiments using Lennard-Jones and Morse potentials. However, there is a principle question: ``Can these bushes exist in real physical systems?'' Indeed, during nonlinear vibrations of real atoms their electron shells polarized and it is not obvious that we can describe the complex physical process of such polarization (characterized by many degrees of freedom!) by simple mechanical mass-point models with pair potential interactions. There is a necessity to verify validity of the concept of bushes of nonlinear normal modes and our group-theoretical methods, aimed at constructing bushes, using real physical experiments. Unfortunately, such direct experiments, as to the best of our knowledge, cannot be fulfilled with the aid of the present experimental technique. However, we can use ab initio simulations based on the density functional theory for checking and deepening the theory of the bushes of NNMs.

\section{Studying bushes of nonlinear normal modes in $SF_6$ molecule with the aid of the density functional theory}\label{sec:3}

The DFT is based on the very non-trivial theorem proved by Kohn and Hohenberg in~\cite{Kohn-Hohenberg}. Indeed, they proved that many particles wave function $\psi(r_1,r_2,r_3,\ldots, r_N)$, depending on $3N$ arguments and satisfying the stationary Schr\"{o}dinger equation for a quantum system with $N$ electrons, can be \emph{exactly} expressed via the function of total electron density, $\rho(\textbf{r})$, depending on \emph{only three} arguments $\textbf{r}=(x,y,z)$. The function $\rho(\textbf{r})$ provides a minimum to a certain functional, $J[\rho(\textbf{r})]$, whose explicit form is not known yet, but one can be sure that it exists in principle. Some approximations were developed for this functional which turn out to be effective and rather accurate.

In our work, we have used ABINIT code developed in~\cite{ABINIT} which realizes computational methods of DFT. This software package provides many useful facilities for ab initio calculations of the structure and different physical properties of molecules, crystals and nanoclusters.

Studying nonlinear vibrations of $SF_6$ molecule, we have used Born-Oppenheimer approximation for separating movement of heavy nuclei and light electrons, as well as local density approximation (LDA) and pseudopotential approach to single out core and outer electrons of the atoms consisting $SF_6$ molecule.

Quantum-mechanical equations by Kohn and Sham~\cite{Kohn-Sham} are used in the ABINIT code for describing dynamics of electrons, while classical equations are solved for nuclei with forces generated by electron shells at any time-step. For solving Kohn-Sham equations the basis of plane waves was used with maximal energy determined by cutoff energy $E_{\rm Cutoff} = 40$~Hartree.

Obviously, one can apply more complicated approximations such as GGA, etc. However, this paper is not aimed at calculations of physical properties which can be verified by real experiments (at the present time, we don't know such experiments). We aimed at confirmation of the general theory of the bushes of nonlinear normal modes in the systems with discrete symmetry using the molecule $SF_6$ as a simple example. In fact, description of this molecule in the framework of the density functional theory with some pertinent approximations represents a mathematical model sufficiently closed to the real physical system.

Let us discuss some numerical results on verification of the bush theory.
\subsection{The one-dimensional bush $B[O_h]$}

First of all, we have calculated equilibrium configuration of $SF_6$ molecule and have found that the edges of the regular octahedron is $a_0=2.9828$ Bohr. This value seems to be rather accurate since experimental value is equal to $2.9554$ Bohr.

\begin{figure}[htb]
\begin{center}
\includegraphics[scale=0.6]{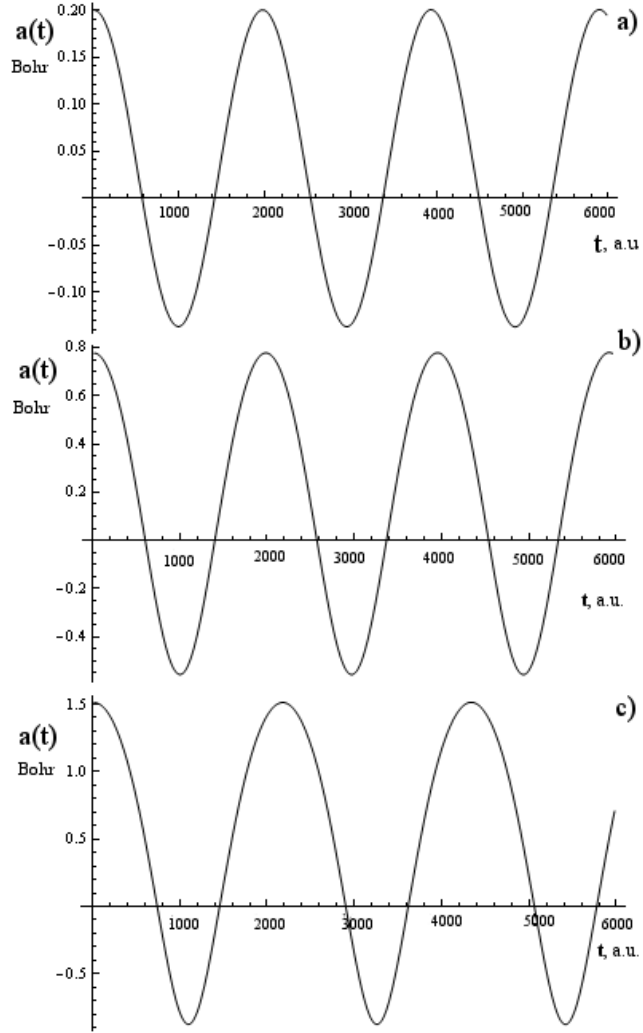}
\caption{Vibrations of the bush $B[O_{h}]$ with different amplitudes. Note that the scale on the vertical axis is different for the cases a, b, c. In all figures, the atomic displacement are given in $Bohrs$, while time $t$ is given in the atomic units ($a.u.$).}\label{fig3}.
\end{center}
\end{figure}

At the next step, we excite vibrations of fluorine atoms by displacing their nuclei by some values, according to the displacement pattern of the breathing mode $\boldsymbol{\phi}_1$ which represents the bush $B[O_h]$ in Table~\ref{table1} (see also Fig. 2a). In other words, we choose the initial shape of the regular octahedron, which can be determined by the edge $a$, larger [$a(0)>a_0$] or smaller [$a(0)<a_0$] than the equilibrium size $a_0$. Then we release the system, i.e.\ permit the molecule to evolve freely. In Fig.~\ref{fig3}, we present the edge $a(t)$ as a function of time for some initial values $a(0)$. From this figure one can see \emph{periodic} vibrations which turns out to be stable, at least, up to the time for which we have observed this dynamical process. The decomposition~\eqref{raz-1} of the corresponding vector $\textbf{X}(t)=[x_1(t), x_2(t),\ldots, x_{18}(t)]$ \emph{does not reveal} (up to the numerical accuracy) any contribution of other modes besides the initially excited breathing mode $\boldsymbol{\phi}_1$. This is in full agreement with the group-theoretical analysis in the previous section.

\begin{figure}[htb]
\begin{center}
\includegraphics[scale=0.8]{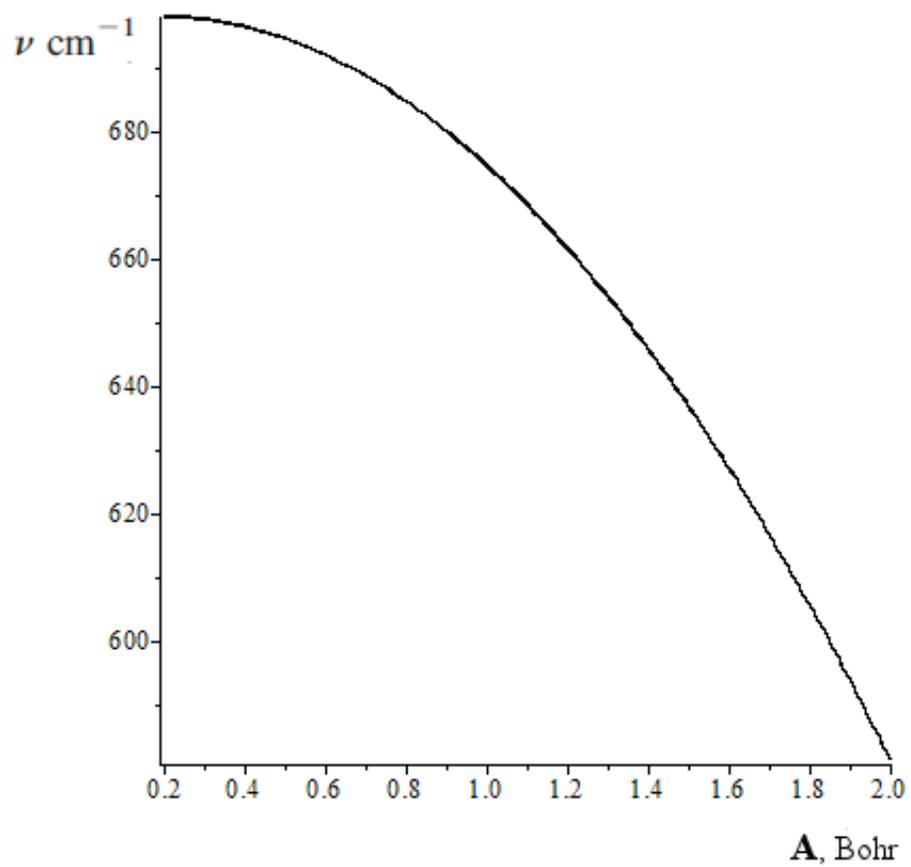}
\caption{Dependence of the frequency $\nu$ of the breathing mode on its amplitude $A$.}\label{fig4}
\end{center}
\end{figure}

In Fig.~\ref{fig4}, we depict the function $\nu(A)$, where $\nu$ is the frequency of the breathing mode and $A$ is its amplitude. As one can see from Fig.~\ref{fig4} the soft nonlinearity realizes in the breathing mode of $SF_6$ molecule.

\subsection{Two-dimensional bush $B[D_{4h}]$}

This bush can be excited by assigning certain values to its both modes at $t=0$: $a_2(0)=\mu$, $a_1(0)=\nu$ with zero initial velocities $[\dot{a_1}(0)=0, \dot{a_2}(0)=0]$. This bush can appear only if $a_2(0)\neq0$ (otherwise the one-dimensional bush $B[O_h]$ will be excited). Indeed, the symmetry of $\boldsymbol{\phi}_2$ is lower than that of $\boldsymbol{\phi}_1$ and, therefore, namely the mode $\boldsymbol{\phi}_2$ turns out to be the root mode for the two-dimensional bush $B[D_{4h}]$.

The most interesting way for the excitation of the bush $B[D_{4h}]$ is the case $\mu\neq 0$, $\nu=0$. In this way, we excite only the root mode $\boldsymbol{\phi}_2$ and have to see the excitation of the secondary mode $\boldsymbol{\phi}_1$, because of the ``force interaction''~\cite{PHD-1998} with the mode $\boldsymbol{\phi}_2$, while all other vibrational modes of $SF_6$ molecule must not be excited.

\begin{figure}[htb]
\begin{center}
\includegraphics[scale=0.55]{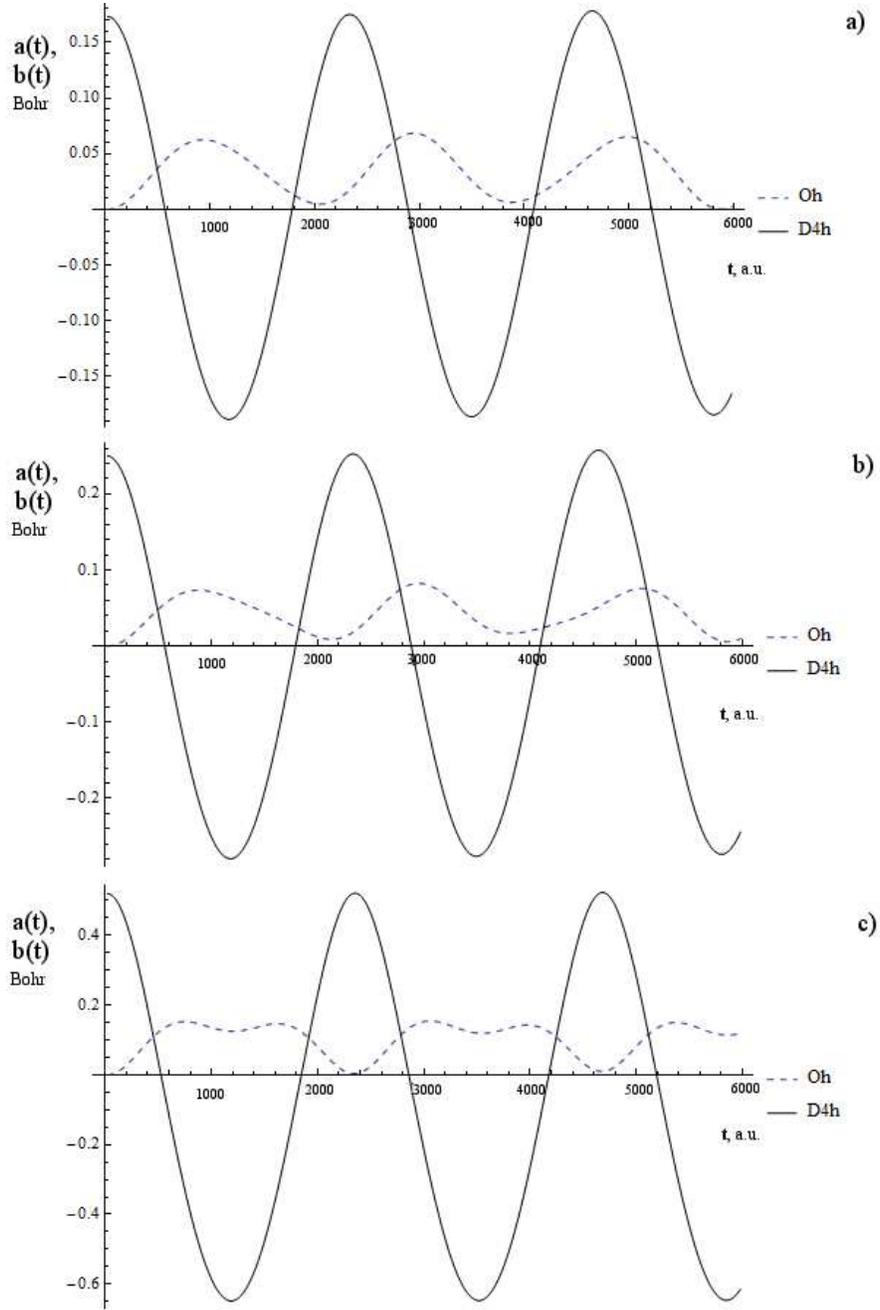}
\caption{Dynamics of the bush $B[D_{4h}]$ with different amplitudes of the root mode $b(t)$.}\label{fig5}
\end{center}
\end{figure}

In Fig.~\ref{fig5}, we present results of such excitation of the bush $B[D_{4h}]$. From this figure one can see that the amplitude $a_1(t)\equiv a(t)$ of the mode $\boldsymbol{\phi}_1$, being zero at $t=0$, is gradually involved into the vibrational process. Then, with time evolution, we see how $a_1(t)$ acts on $a_2(t)$ and vice versa, or interaction between modes $\boldsymbol{\phi}_1$ and $\boldsymbol{\phi}_2$.

\begin{figure}[htb]
\begin{center}
\includegraphics[scale=0.8]{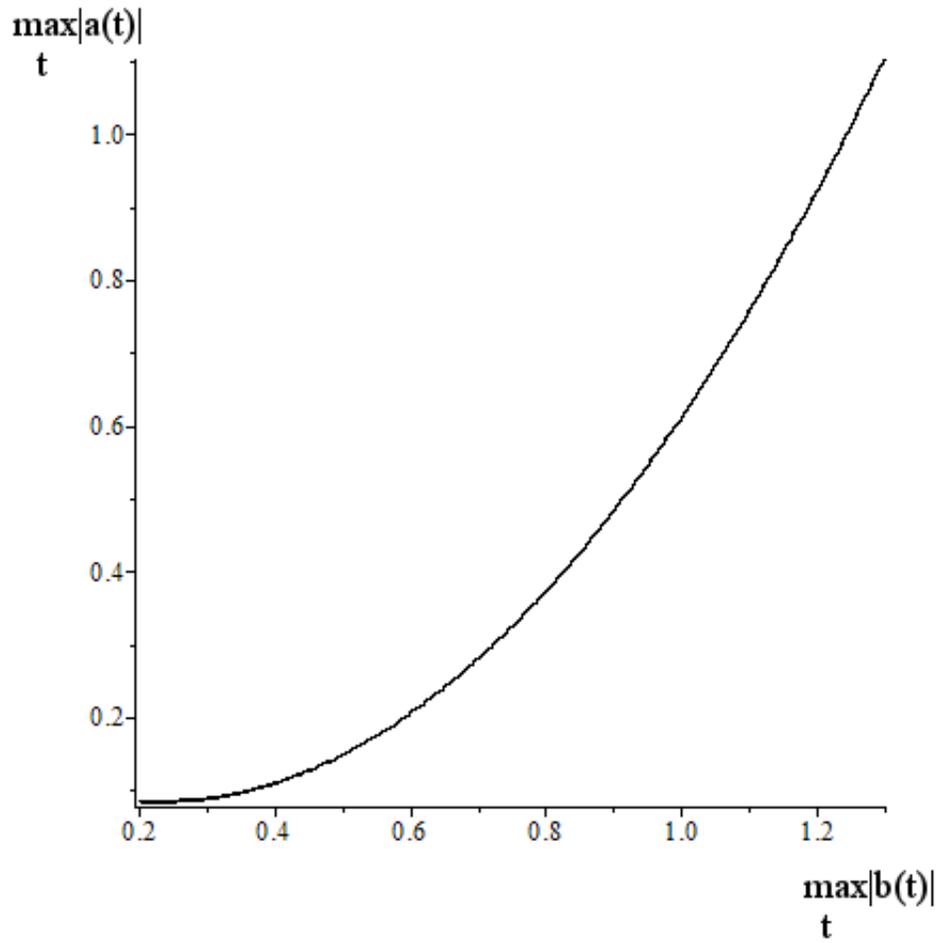}
\caption{Dependence of $\max|a_1(t)|$ on $\max|a_2(t)|$.}\label{fig6}
\end{center}
\end{figure}

Fig.~\ref{fig6} illustrates the dependence of $\underset{t}{\max}|a_1(t)|$ on $\underset{t}{\max}|a_2(t)|$. Such dependence turns out to be important if we want to study the excitation of the secondary mode $a_1(t)\boldsymbol{\phi}_1$ with the aid of the perturbation theory.

Every bush can be treated as a closed Hamiltonian system. In the present case of the bush $D_{4h}$, we have such a system described by two dynamical variables $a_1(t)=a(t)$, $a_2(t)=b(t)$. In~\cite{PHD-1998, DAN-2}, we study the ``classes of dynamical universality'' of the bushes. This concept appears naturally if one studies the decomposition of potential energy of different bushes into multidimensional Taylor series. Indeed, it turns out that many bushes in different physical systems, and even in the same system possess \emph{identical forms} of the potential energy decomposition up to a certain degree. Because of this reason it is interesting to analyse decomposition of the potential energy of the bush $B[D_{4h}]$. This can be done as follows.

Let us choose a certain grid in the space of two variables $\mu$ and $\nu$. Each node of this grid determines the values of dynamical variables $a(t)$ and $b(t)$ of the bush $B[D_{4h}]$ at $t=0$. The linear combination of the mode vectors $\boldsymbol{\phi}_1$, $\boldsymbol{\phi}_2$ with the coefficients $\mu_{ij}$, $\nu_{ij}$, corresponding to the chosen ($i,j$) node, determines a certain displacement pattern of the fluorine atoms
\begin{equation}\label{9}
\textbf{X}_{ij}(0)=\mu_{ij}\boldsymbol{\phi}_1+\nu_{ij}\boldsymbol{\phi}_2
\end{equation}
and, therefore, a definite initial configuration of the $SF_6$ molecule.

On the other hand, using ABINIT code we can find the potential energy $u_{ij}$ of this configuration [for each node this program finds the electron density corresponding to the positions of the fluorine atoms determined by~\eqref{9}]. Then one can construct the potential energy $U(a,b)$ as a polynomial of a fixed degree in variables $a, b$:
\begin{equation}
U(a,b)=\sum_{m,n}\gamma_{mn}a^m b^n.
\end{equation}
We determine coefficients $\gamma_{mn}$ by the least square method. As a result, the following potential energy $U(a,b)$ of the bush $B[D_{4h}]$ was obtained:
\begin{equation}\label{U_d4h}
\begin{split}
U(a,b)&=0.13418b^2+0.19269a^2-0.20382ab^2+0.04220b^3-0.09507a^3\\
&\quad -0.04813ab^3+0.12396a^2b^2+0.03085b^4+0.03013a^4.
\end{split}
\end{equation}
Here we point out only those terms of $U(a,b)$ which are permitted by the symmetry group $D_{4h}$, i.e.\ Eq.~\eqref{U_d4h} represents decomposition of $U(a,b)$ into the first \emph{polynomial invariants}~\cite{PHD-1998} of the group $G=D_{4h}$.

Constructing $U(a,b)$ with the aid of the least square method, we have also calculated polynomial terms which are not permitted by the group $D_{4h}$ to verify that they are rather small (about $10^{-3}-10^{-4}$).

Note that the similar expression for $U(a, b)$ we have already obtained in~\cite{octo} applying the Taylor expansion of the exact potential energy for octahedral mechanical structure whose particles interact via Lennard-Jones potential.

Now, we can write Newton equations for the two-dimensional system corresponding to the bush $B[D_{4h}]$:
\begin{equation}\label{Eq_d4h}
\begin{split}
\ddot{a}+0.38538a&=0.28521a^2+0.20382b^2-0.24791ab^2-0.12051a^3+0.04813b^3,\\
\ddot{b}+0.26835b&=b(-0.12658b+0.40763a-0.12340b^2-0.24791a^2+0.14440ab).
\end{split}
\end{equation}

From these equations, one can see ``inequality of rights'' of the root and secondary modes. Indeed, if we excite at $t=0$ \emph{only secondary mode} $a(t)$ by appropriate initial conditions $[a(0)\neq 0$, $\dot{a}(0)=b(0)=\dot{b}(0)=0]$, it will vibrate for arbitrary long time according to the equation $\ddot{a}+0.38538a=0.28521a^2-0.12051a^3$, while the root mode remains to be unexcited: $b(t)\equiv 0$. In the contrary case, where \emph{only root mode} $b(t)$ is excited by the initial conditions $[b(0)\neq 0$, $\dot{b}(0)=a(0)=\dot{a}(0)=0]$, the secondary mode, $a(t)$, cannot be equal to zero, because there are ``external'' forces proportional to $b^2(t)$ and $b^3(t)$ in the r.h.s.\ of the first equation~\eqref{Eq_d4h}.

According to the bush theory (see~\cite{PHD-1998}) this fact is a straightforward consequence of the difference in symmetries of the modes $a(t)\boldsymbol{\phi}_1$ and $b(t)\boldsymbol{\phi}_2$ which are $O_h$ and $D_{4h}$, respectively (the group $D_{4h}$ is the \emph{subgroup} of the group $O_h$).

\begin{figure}[htb]
\begin{center}
\includegraphics[scale=0.55]{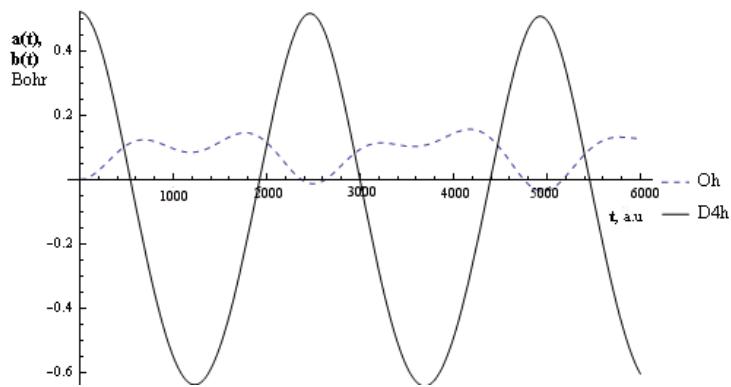}
\caption{Approximate dynamics of the bush $B[D_{4h}]$.}\label{fig7}
\end{center}
\end{figure}

Solving Eqs.~\eqref{Eq_d4h} numerically with the aid of fourth order Runge-Kutta method we obtain some examples of the bush $D_{4h}$ dynamics. Such example is presented in Fig.~\ref{fig7}. One ought to compare this figure with the similar Fig.~\ref{fig5} which has been obtained by the ABINIT code. The comparison shows that Eqs.~\eqref{Eq_d4h}, describing dynamics of the bush $B[D_{4h}]$ as an autonomous two-dimensional system, turn out to be rather correct for the considered vibrational amplitudes.

\subsection{Three-dimensional bush $B[C_{4v}]$}

According to Eq.~\eqref{raz-c4v}, the three-dimensional bush $B[C_{4v}]$ can be written in the form
\begin{equation}\label{50}
\textbf{X}(t)[C_{4v}]=a(t)\boldsymbol{\phi}_1+b(t)\boldsymbol{\phi}_2+c(t)\boldsymbol{\phi}_3.
\end{equation}
Here, $\boldsymbol{\phi}_1$, $\boldsymbol{\phi}_2$, $\boldsymbol{\phi}_3$ are the vibrational modes from Table 1 which determine the special displacement patterns of the fluorine atoms of the molecule $SF_6$, while $a(t)$, $b(t)$, $c(t)$ are time-dependent functions describing bush evolution in time.

As was already discussed in Sec.~\ref{sec:2}, the symmetry groups of the modes $\boldsymbol{\phi}_1$, $\boldsymbol{\phi}_2$, $\boldsymbol{\phi}_3$ are $O_h$, $D_{4h}$, $C_{4v}$, respectively. The following group-subgroup relation between these point groups exists:
\begin{equation}\label{relation}
O_h\supset D_{4h}\supset C_{4v}.
\end{equation}
Thus, the mode $\boldsymbol{\phi}_3$ has the lowest symmetry group among three modes of the bush $C_{4v}$ and, therefore, it turns out to be root mode of this bush. The point group of the whole bush $B[C_{4v}]$ coincides with that of the mode $\boldsymbol{\phi}_3$. The modes $\boldsymbol{\phi}_2$ and $\boldsymbol{\phi}_1$ are secondary modes of the considered bush, and according to the bush theory~\cite{PHD-1998} they must be automatically involved into the vibrational process as a consequence of the initial excitation of the root mode $\boldsymbol{\phi}_3$.

To verify this prediction we have fulfilled simulation of $SF_6$ molecule vibrations with the aid of ABINIT code. Results of these simulations are represented in Fig.~\ref{fig8} for three different values of the initial amplitudes $c(0)$ of the root mode $\boldsymbol{\phi}_3$ (two other initial amplitudes, as well as velocities of all three modes of the bush $B[C_{4v}]$ are assumed to be zero). This figure illustrates the excitation of the secondary modes $\boldsymbol{\phi}_2$ and $\boldsymbol{\phi}_3$ because of their interactions with the root mode $\boldsymbol{\phi}_3$.

Using the least square method, as it was described for the bush $B[D_{4h}]$, we have obtained the following polynomial expression for the potential energy $U(a,b,c)$ of the bush $B[C_{4v}]$ as the function of three variables $a, b, c$:
\begin{equation}\label{U_c4v}
\begin{split}
U(a,b,c)&=0.19280a^2+0.15057c^2+0.13254b^2+0.21180bc^2-0.23103ac^2-0.22387ab^2\\
&\quad +0.04395b^3-0.08708a^3+0.02771794903c^4+0.16522b^2c^2+0.03186b^4\\
&\quad -0.2562abc^2-0.05031ab^3+0.13336a^2c^2+0.13581a^2b^2+0.02575a^4.
\end{split}
\end{equation}

Note that only polynomial \emph{invariants} of the group $G_3=C_{4v}$ give large contributions to $U(a,b,c)$ (all other polynomial terms obtained by direct least square method turn out to be rather small).

It is very interesting to note that according to the theorems proved in~\cite{PHD-1998}, group-subgroup relation~\eqref{relation} provides the special structure of the bush $B[C_{4v}]$ dynamical equations. Indeed, the potential energy $U(a,b,c)$ contains invariants which are \emph{linear} in secondary modes, while they contain some degrees of root mode. We see such invariants in Eq.~\eqref{U_c4v}:
\begin{equation}\label{invariants}
c^2a, c^2b.
\end{equation}

The Newton dynamical equations for the bush $B[C_{4v}]$ as a closed Hamilton system can be written as follows:
\begin{equation}\label{eq_mot_c4v}
\begin{split}
   \ddot{a}+0.3856a&=0.26125a^2+0.22387b^2+0.23103c^2-0.10301a^3\\
   &\quad +0.05031b^3-0.27163ab^2-0.26672ac^2+0.256bc^2,\\
   \ddot{b}+0.26508b&=-0.13184b^2-0.21180c^2+0.44775ab-0.12743b^3\\
   &\quad -0.27163a^2b+0.15093ab^2+0.256ac^2-0.33045bc^2,\\
   \ddot{c}+0.30114c&=c(0.46205a-0.42361b-0.11087c^2-0.26672a^2\\
   &\quad +0.512ab-0.33045b^2).\\
\end{split}
\end{equation}

The invariants~\eqref{invariants} produce in the right-hand side of Eqs.~\eqref{eq_mot_c4v} \emph{forces} which act from the root mode $c(t)$ on the secondary modes $a(t)$ and $b(t)$. Namely, these forces form the bush as unique dynamical object. The full set of its modes conserves in time, while their amplitudes $a(t)$, $b(t)$, $c(t)$ do change.

\subsection{Nonlinearity in $SF_6$ dynamics}

In this paper, essentially nonlinear vibrations of $SF_6$ molecule are considered. Unfortunately, we do not know any experimental data that allow a direct comparison with our results. However, in the case of oscillations with small amplitudes these results should be consistent with those of the linear normal modes analysis. Let us discuss this point in more detail.

Obtained in the harmonic approximation, linear normal modes represent sinusoidal oscillations whose frequencies do not depend on their amplitudes, and all such modes are independent of each other. This means that if only one LNM was excited in the considered system, then it will exist for arbitrary long time without excitation of any other normal mode.

Results of our ab initio simulations demonstrate a violation of these properties for oscillations with large amplitudes. It can be seen from Fig.~\ref{fig3} that oscillations of the mode $a(t)\boldsymbol{\phi}_1$ with symmetry group $O_h$ are not strictly sinusoidal. The frequency $\nu$ of this mode depends on its amplitude $A$, and we depict the dependence $\nu(A)$ in Fig.~\ref{fig4}. For small amplitudes the frequency changes slightly, while for large amplitudes the considerable decreasing of $\nu(A)$ with increasing of the amplitude takes place. Therefore, $a(t)\boldsymbol{\phi}_1$ vibrations demonstrate soft type of nonlinearity. In the limit $A\rightarrow 0$, the frequency of the discussed mode tends approximately to $710~\text{cm}^{-1}$.This value may be compared with the frequencies $718~\mathrm{cm}^{-1}$ \cite{38} and $723.5~\text{cm}^{-1}$~\cite{bruska} for the mode $\nu_1[A_g]$, which were found with the aid of $DMol_3$ code, and with the experimental value $772.27~\text{cm}^{-1}$. Possible reasons for discrepancy between above given values of $\nu_1[A_g]$ will be analysed elsewhere. In this paper, we focus only on confirmation of the general bush theory of nonlinear vibrations in physical systems with discrete symmetry using $SF_6$ molecule as a simple example. According to this theory in the case of large amplitudes, we should observe a very definite interaction between normal modes which were negligible for the small-amplitude vibrations.

In Fig.~\ref{fig5}, we see interaction between the vibrational modes $b(t)\boldsymbol{\phi}_2$ and $a(t)\boldsymbol{\phi}_1$ whose symmetries are $D_{4h}$ and $O_h$, respectively. Indeed, at the initial instant $t=0$ only mode $b(t)\boldsymbol{\phi}_2$ (root mode) was excited. However, the excitation from this mode is gradually transferred to the mode $a(t)\boldsymbol{\phi}_1$ which is the secondary mode [its symmetry is higher than that of the mode $b(t)\boldsymbol{\phi}_2$]. The interaction between the above modes is very important since the amplitude of the secondary mode reaches $30\%$, and even more, as compared with that of the root mode. Moreover,
dynamics of the secondary mode is radically different from the simple sinusoidal oscillations, and this difference becomes more visible with increasing amplitude of the root mode.

The main point is that excitation from the root mode is transferred to only one of all other modes, and exactly to the mode which must be excited in accordance with the bush theory (excitation can be transferred only to the modes with higher or equal symmetry).	

In our case, the secondary mode $a(t)\boldsymbol{\phi}_1$ is associated with the two-dimensional irrep with conventional optics symbol $E_g$. We have obtained the frequency $\nu[D_{4h}]$ of this mode for small-amplitude vibrations with $9\%$ error according to its experimental value, while the root-mode frequency $\nu[O_h]$ was calculated with $8\%$ error. However, it is interesting to note that the ratio $\nu[O_h]\setminus\nu[D_{4h}]$ found in our calculations is equal to $1.2$, and this value almost exactly coincides with that obtained from the experimental data.

\begin{figure}[htb]
\begin{center}
\includegraphics[scale=0.55]{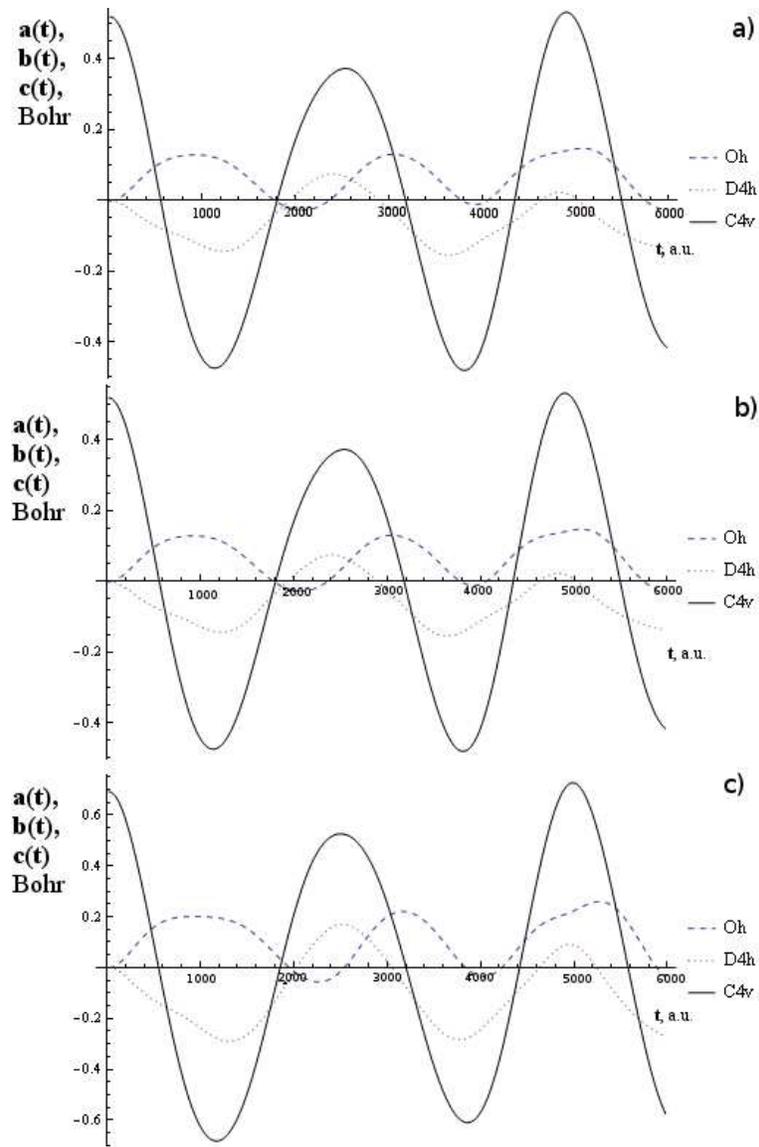}
\caption{Vibrations of the bush $B[C_{4v}]$ with different amplitudes of the root mode $c(t)$.}\label{fig8}
\end{center}
\end{figure}

All above discussed manifestations of nonlinearity in dynamics of the two-dimensional bush $B[D_{4h}]$ can be also found in time-evolution of the three-dimensional bush $B[C_{4v}]$ whose dynamics is shown in Fig.~\ref{fig8}. From this figure one can see that two secondary modes $a(t)\boldsymbol{\phi}_1$ and $b(t)\boldsymbol{\phi}_2$ only these modes!) are involved into vibrational regime due to the interaction with the root mode $c(t)\boldsymbol{\phi}_3$ associated with the three-dimensional irrep~$F^1_u$.

According to~\cite{bruska}, the infrared absorption spectrum of $SF_6$ is composed of two bands whose maxima are located at $\nu_3=938~\text{cm}^{-1}$ and $\nu_4=567~\text{cm}^{-1}$. Both of these normal modes are triple degenerate since they belong to the irrep $F^1_u$. Our rough estimate of $c(t)\boldsymbol{\phi}_3$-mode frequency in the case of small amplitude vibrations gives $\nu[C_{4v}]=530~\text{cm}^{-1}$ that allows us to identify this frequency with $\nu_4$.

The Figs.~\ref{fig5} and~\ref{fig8} clearly show that the bush with dimension $m>1$ represents a quasi-periodic dynamical object because it is a superposition of modes with different frequencies.

In conclusion, let us note that only one mode, $c(t)\boldsymbol{\phi}_3$, belonging to the three-dimensional irrep $F^1_u$ contributes to the bush $B[C_{4v}]$. What would happen if we initially excite another mode of the irrep $F^1_u$? It is easy to understand that the result of this action will be emergence of another bush which turns out to be \emph{dynamically equivalent} to the bush $B[C_{4v}]$. Such bushes we call ``dynamical domains'' of one and the same bush, in full accordance with the corresponding term of the theory of phase transitions in crystals.

In the case of the irrep $F^1_u$ there exist three dynamical domains of the bush $B[C_{4v}]$ which differ only in directions of the atomic vibrations. The axis of symmetry of the above examined bush $B[C_{4v}]$ coincides with the coordinate axis $Z$, while the axes of symmetry of two other its domains coincide with $X$ and $Y$ coordinate axes.

\section{Conclusion}
The main goal of this paper was to verify the validity of the group-theoretical results obtained by the theory of bushes of nonlinear normal modes in physical systems with discrete symmetry using as an example $SF_6$ molecule. We have also investigated some properties of one-, two- and three-modes exact nonlinear vibrational regimes (bushes of NNMs) in this molecule with the aid of ab initial simulations based on the density function theory. For these calculations we have used ABINIT code~\cite{ABINIT}.

Some problems concerning the above bushes in $SF_6$ molecule turn out to be beyond the scope of the present paper. They will be studied in further works. Some of these problems are:

1. Stability of the bushes of NNMs in $SF_6$ molecule which were obtained by the ABINIT code;

2. The method for exciting these bushes by external fields;

3. The possibility to describe large nonlinear vibrations in $SF_6$ in the framework of mass-point models whose particles interact via pair phenomenological potentials;

4. The possibility of applying the theory of bushes of nonlinear normal modes for studying some structural phase transitions whose nature relates to vibration of octahedron clusters in perovskite-like crystals.

\section*{Acknowledgements}

The work was partly supported by the Russian Science
Foundation (Grant No.\ 14-13-00982). The present results
have been obtained through the use of the ABINIT code,
a common project of the Universitie Catholique de Louvain,
Corning Incorporated, and other contributors (URL
\url{http://www.abinit.org}).

\section*{Figures}


\end{document}